# Ultra-high Q lithium niobate microring monolithically fabricated by photolithography assisted chemo-mechanical etching

CHUNTAO LI,[1,2] JIANGLIN GUAN,[1,2] JINTIAN LIN[3,5,†], RENHONG GAO,[3,5] MIN WANG,[1,2,4] LINGLING QIAO,[3] LI DENG,[1,2] AND YA CHENG[1,2,3,5,6,7,8,9,*]

[1] *State Key Laboratory of Precision Spectroscopy, East China Normal University, Shanghai 200062, China*
[2] *The Extreme Optoelectromechanics Laboratory (XXL), School of Physics and Electronic Science, East China Normal University, Shanghai 200241, China*
[3] *State Key Laboratory of High Field Laser Physics and CAS Center for Excellence in Ultra-Intense Laser Science, Shanghai Institute of Optics and Fine Mechanics (SIOM), Chinese Academy of Sciences (CAS), Shanghai 201800, China*
[4] *Engineering Research Center for Nanophotonics & Advanced Instrument, Ministry of Education, School of Physics and Electronic Science, East China Normal University, Shanghai 200241, China*
[5] *Center of Materials Science and Optoelectronics Engineering, University of Chinese Academy of Sciences, Beijing 100049, China*
[6] *Collaborative Innovation Center of Extreme Optics, Shanxi University, Taiyuan 030006, China*
[7] *Collaborative Innovation Center of Light Manipulations and Applications, Shandong Normal University, Jinan 250358, China*
[8] *Shanghai Research Center for Quantum Sciences, Shanghai 201315, China*
[9] *Joint Research Center of Light Manipulation Science and Photonic Integrated Chip, East China Normal University, Shanghai 200241, China*

[†]*jintianlin@siom.ac.cn*
[*]*ya.cheng@siom.ac.cn*

Date: 18 June 2023

**Abstract:** Thin-film lithium niobate (TFLN) has been considered as one of the most important platforms for constructing high-performance photonic integrated devices such as electro-optic modulators, frequency combs, classical/quantum light sources, and large-scale photonic integrated circuits, benefiting from its excellent optical properties of TFLN. The fabrication quality of TFLN photonic integrated devices plays an important role in the performance and the integration scale of these devices. As one of the element photonic structures, the state-of-the-art TFLN microrings reach an intrinsic Q factor higher than $10^7$ with ultra-smooth sidewalls, fabricated by photolithography assisted chemo-mechanical etching (PLACE). However, it is isolated on the chip surface and a tapered fiber is required to couple the light into and out of the resonator. Furthermore, it is difficult to maintain such high-Q factors when the microrings are monolithically integrated with bus waveguides by PLACE, resulted from large coupling loss with biggish coupling gap. Here, a relatively narrow gap of an ultra-high Q microring monolithically integrated with the bus-waveguide is achieved with 3.8 μm by optimizing PLACE process, and a high temperature annealing is carried out to improve the loaded (intrinsic) Q factor with $4.29\times10^6$ ($4.04\times10^7$), leading an ultra-low propagation loss of less than 1 dB/m, which is approximately 3 times better than the best values previously reported in ion-slicing TFLN platform.

## 1. Introduction

Thin-film lithium niobate (TFLN) platform has shown great potentials for building of large-scale photonic integrated circuits (PICs) and is highly favored in fundamental research, because of the excellent properties of TFLN such as the broad transparency windows with high refractive index contrast between conventional cladding materials, high linear electro-optic coefficient and second-order nonlinear optical coefficient, acousto-optic effect, and piezo effect [1-17]. For lots of PICs applications, high performance, low power consumption, and scalable large-scale integration, are the core figures of merits, which are high sensitivity to the propagation loss of the devices. The propagation loss is mainly contributed from the surface roughness, coupling loss, and inner scattering loss of the devices [1-17]. To suppress the surface roughness, photolithography assisted chemo-mechanical polishing has been used to etch the TFLN, leading to ultra-smooth sidewall [18-23]. So far, ultra-high free-standing TFLN microdisk resonators and meter-scale length low-loss waveguides have been demonstrated this novel platform [18,19,24], opening up the new possibilities of their applications ranging from low-threshold nonlinear optical processes, microlasers to high-speed optical information processing. However, due to the limited etch ratio between chromium hard mask and lithium niobate [18-24], photonic structures are often produced with low aspect ratio, making obstacle for monolithic microring side-coupled with bus-waveguide. Whereas efforts have been made towards a monolithic integration scheme, that is, where a microring was over-coupled with the bus ridge-waveguide on the same chip, and the optical modes in the microring and the waveguide could be bridged with a slab lithium niobate layer to couple with each other [21]. Up to now, the narrowest coupling gap between microring and the bus-waveguide has been reduced to 4.8 μm with a slab layer with thickness of 500 nm and an etched depth of 200 nm. The over-coupled condition introduces high coupled loss and relatively low coupling efficiency of ~10%, leading to a relatively low Q loaded factors of $3.2\times10^5$, which is much lower than the isolated counterpart. Moreover, the light confinement in the rib waveguide is weaker than the strip counterpart, which will make an obstruction to nonlinear optical applications and high-density integration.

Here, a monolithic microring side-coupled with rib waveguide was demonstrated on single TFLN chip with high loaded Q factor. The gap of the coupling region was bridged to 3.8 μm to obtain a higher coupling efficiency (37%) by optimizing the PLACE process. To further improve the Q factors, high temperature annealing was adopted to restore the lattice damage caused by ion implantation during ion-slicing step, leading to an improvement in the loaded Q factors with 4 times. The highest loaded Q factors reach $4.29\times10^6$, which is of the same order of magnitude as those of the best microrings fabricated using ICP-RIE (inductively coupled plasma - reactive ion etching) [24-29]. And the intrinsic Q factor was determined as high as $4.04\times10^7$, corresponding to a propagation loss of 0.0091 dB cm$^{-1}$, which is approximately 3 times lower than the best results reported on ion-slicing TFLN platform [23,24,28]. Our work will pave the way for the high-performance classical/quantum nonlinear light sources, optical information processing, and large-scale PICs with high scalability.

## 2. Fabrication Methods

In our experiment, the monolithically integrated on-chip microring resonator side-coupled with a strip waveguide is fabricated on an undoped TFLN wafer, as shown in Fig. 1(a), which is composed of 700 nm-thick Z-cut TFLN, 2000 nm-thick silicon dioxide (SiO$_2$) layer, and 500 μm-thick lithium niobate substrate.

The manufacturing process to fabricate the microring resonator comprises five major steps, as illustrated in Fig. 1(b)-(f). First, a 600 nm-thick chromium (Cr) film is coated on the surface of the thin-film lithium niobate on insulator (LNOI) by magnetron sputtering. Second, the Cr layer is ablated into the hard mask with pattern consisted with microrings side-coupled with strip waveguide by femtosecond laser direct writing [18]. In this process, the curvature radius of the microring is set to be 200 μm with a top width of 2.35 μm and the coupling gap between microring and strip waveguide is designed to be 3.8 μm. Third, the fabricated structure undergoes chemo-mechanical polishing (CMP) to etch the exposed TFLN, leading to the pattern transferring from the Cr layer to TFLN. Fourth, the sample is immersed in the Cr etching solution to remove the Cr mask. Finally, a secondary CMP and a high temperature annealing (450°C for 2 hours in air) are carried out to restore the lattice damage to guarantee a high Q factor of the microring resonator.

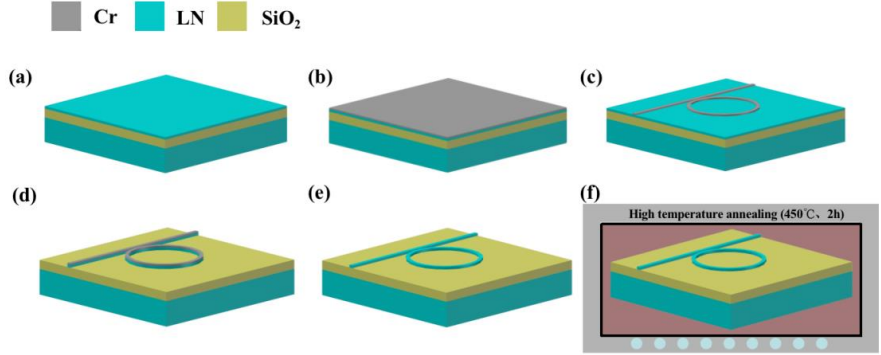

Fig. 1. Illustration of the fabrication flow of the monolithically integrated microring resonator. (a) Sandwiched structure configuration of an undoped 700 nm-thick Z-cut TFLN wafer. (b) The Cr film is coated on the surface of the undoped LN wafer. (c) Cr mask pattern is formed by femtosecond laser direct writing. (d) The exposed TFLN is etched by CMP for transferring the Cr mask pattern to the TFLN layer. (e) The Cr mask is removed by chemical wet etching. (f) The sample undergoes a secondly CMP and a high temperature annealing.

## 3. Characteristics of LN Microrings Side Coupled with a Ridge Waveguide

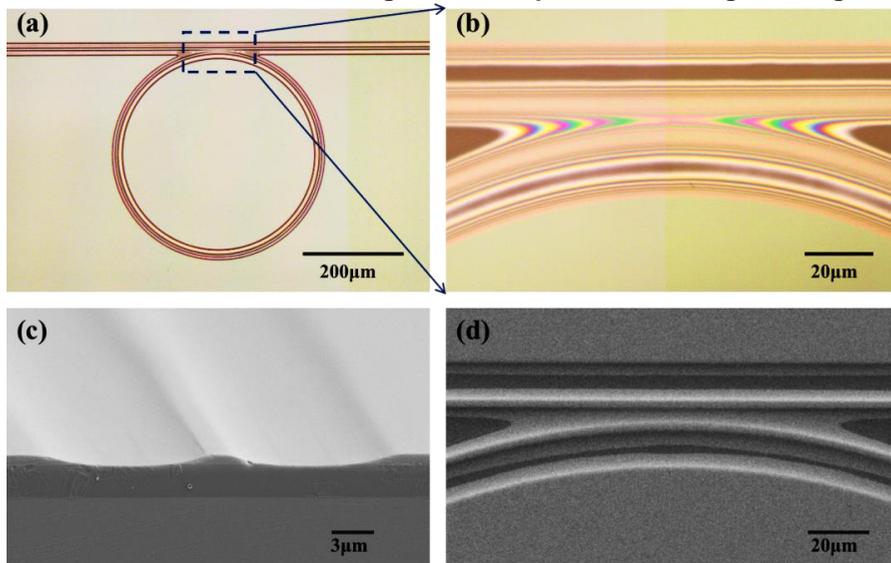

Fig. 2. (a) Optical microscope image of the fabricated microring resonator. (b) The close-up optical micrograph image of the fabricated microring resonator. (c) The scanning-electron-microscope image of the cross-section of the strip waveguide. (d) The magnified SEM image of the coupling region between the microring resonator and the straight waveguide.

The optical microscope image of the fabricated on-chip microring resonator is shown in Fig. 2(a). The zoom-in optical micrograph and the magnified scanning electron-microscope (SEM) image of the coupling region between the microring resonator and the strip waveguide are shown in Figs. 2(b) and 2(d), respectively, indicating the TFLN microring with a diameter of

400 µm and an ultra-smooth surface. The gap between the strip waveguide and the ring resonator was measured to be 3.8 µm. The TFLN waveguides with a top width of 2.35 µm, bottom width of 9.68 µm and a thickness of 700 nm are located on the top of a 2 µm-thick $SiO_2$ layer. The fabricated LN waveguide is slightly thinner than the original TFLN layer due to the in Figure 2(c). The top width of the microring and the strip waveguide was 1.75 µm. The thickness of the microring has been thinned to be 645.4 nm, and the wedge angle of the waveguide was 10°.

## 4. Results and Discussion

We characterized the Q factors of the modes of the fabricated integrated microring resonator before and after the annealing by sweeping the laser wavelength at 1550 nm wave band and measuring the transmission spectra. Lensed fibers were used to couple the light signal into and out of the strip waveguide by end-fire coupling. A C-band wavelength tunable narrow-linewidth laser diodes (model: TLB-6728, New Focus Inc.) was used as pump laser to measure the loaded Q factor. The power coupled into the microring resonator from the pump source needs to be as low as possible to avoid thermal and other nonlinear optical effects, which is ~ 5 µW. The polarization state of the input light was tuned with an inline polarization controller. The output optical signal was coupled out of the microring resonator by the same strip waveguide and lensed fiber and sent into a photodetector (PD, model: 1611, New Focus Inc.). The transmission spectrum was real-time analyzed by an oscilloscope (model: Tektronix MDO04). We can obtain sharp dips in the transmission spectrum when the laser wavelength was resonant with the cavity modes. And the Q factor of the microring resonator was determined by fitting the resonant dips with Lorentz function.

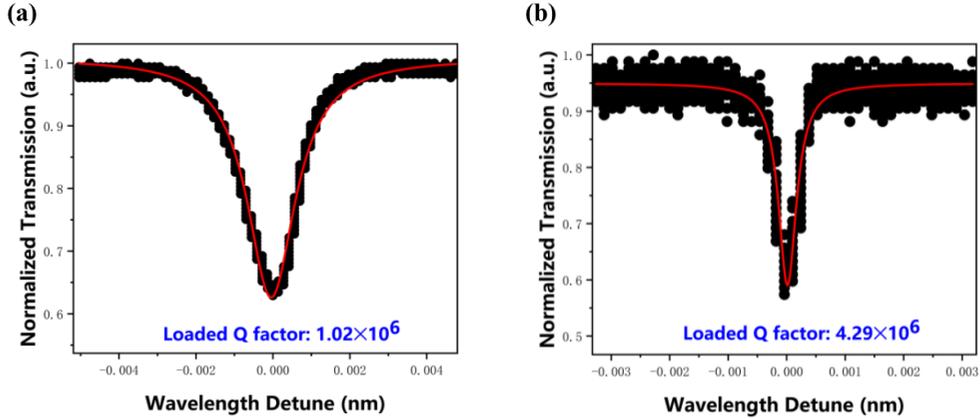

Fig. 3. (a) Transmission spectra before annealing (black dots) and Lorentz fitting (red curve) revealing a loaded Q factor of $1.02\times10^6$ at the resonant wavelength of 1556.6 nm. (b) Transmission spectra after annealing (black dots) and Lorentz fitting (red curve) revealing a loaded Q factor of $4.29\times10^6$.

A typical fundamental mode at 1556.6 nm wavelength is plotted in Fig. 3(a) before annealing, which exhibits a Lorentz-shape curve. The coupling efficiency was measured to be 37%, which is higher than the results fabricated by PLACE technique [18]. And the loaded Q factor reached $1.02\times10^6$, which is resulted from the ultra-smooth sidewall of the photonic device. After annealing, the loaded Q factor up to $4.29\times10^6$ was achieved, as shown in Fig 3(b). It is more than 4 times higher than the result before annealing. And the coupling efficiency remains almost unchanged before and after annealing. Since the strip waveguide is relatively far from the microring, the waveguide is under-coupled with the microring. Therefore, the intrinsic Q factor of the mode was calculated to be $4.04\times10^7$. This high intrinsic Q factor is closed to be the best results previously reported using PLACE technique [18,23].

Furthermore, we systemically measured the loaded Q factors of the same monolithic microring resonators before and after annealing for comparison, as shown in Table 1. The results show that the loaded Q factors are significantly improved with 4-5 times after the high temperature annealing. The underlying physics behind it is that the annealing can restore the lattice damage of the material which was induced by ion implantation when producing the TFLN [30-32].

Table 1. Comparison of the loaded Q factors of the microring resonators before and after annealing

| Resonant wavelength (nm) | Before Annealing | | After Annealing | |
|---|---|---|---|---|
| | Q ($\times 10^6$) | Transmissivity | Q ($\times 10^6$) | Transmissivity |
| 1556.6 | 1.02 | 63% | 4.29 | 62% |
| 1548.2 | 0.91 | 90% | 3.65 | 90% |
| 1560.0 | 0.58 | 90% | 2.27 | 91% |
| 1531.0 | 0.57 | 93% | 2.35 | 93% |
| 1549.0 | 1.01 | 94% | 4.83 | 94% |
| 1524.8 | 0.68 | 95% | 3.86 | 96% |
| 1540.3 | 0.51 | 96% | 2.40 | 96% |

## 5. Conclusion

In conclusion, we have demonstrated monolithically integrated microring with ultra-high Q factor on TFLN platform, which will open up new avenue for applications ranging from classical/quantum integrated light sources to coherent optical communications, metrology, and LiDAR application.